\documentclass[twocolumn,american,english]{revtex4}
\usepackage[T1]{fontenc}
\usepackage[latin9]{inputenc}
\usepackage{amsmath}
\usepackage{graphicx}

\makeatletter

\providecommand{\tabularnewline}{\\}

\@ifundefined{textcolor}{}
{%
 \definecolor{BLACK}{gray}{0}
 \definecolor{WHITE}{gray}{1}
 \definecolor{RED}{rgb}{1,0,0}
 \definecolor{GREEN}{rgb}{0,1,0}
 \definecolor{BLUE}{rgb}{0,0,1}
 \definecolor{CYAN}{cmyk}{1,0,0,0}
 \definecolor{MAGENTA}{cmyk}{0,1,0,0}
 \definecolor{YELLOW}{cmyk}{0,0,1,0}
}

\makeatother

\usepackage{babel}
\begin{document}

\title{Global Method for Electron Correlation}

\author{Mario Piris }

\address{Kimika Fakultatea, Euskal Herriko Unibertsitatea (UPV/EHU), P.K.
1072, 20080 Donostia, Euskadi (Spain);}

\address{Donostia International Physics Center (DIPC), 20018 Donostia, Euskadi
(Spain)}

\address{IKERBASQUE, Basque Foundation for Science, 48013 Bilbao, Euskadi
(Spain).\vspace{0.8cm}
}
\begin{abstract}
The current work presents a new single-reference method for capturing
at the same time the static and dynamic electron correlation. The
starting-point is a determinant wavefunction formed with natural orbitals
obtained from a new interacting-pair model. The latter leads to a
natural orbital functional (NOF) capable of recovering the complete
intra-pair, but only the static inter-pair correlation. Using the
solution of the NOF, two new energy functionals are defined for both
dynamic ($E^{dyn}$) and static ($E^{sta}$) correlation. $E^{dyn}$
is derived from a modified second-order Møller\textendash Plesset
perturbation theory (MP2), while $E^{sta}$ is obtained from the static
component of the new NOF. Double counting is avoided by introducing
the amount of static and dynamic correlation in each orbital as a
function of its occupation. As a result, the total energy is represented
by the sum $\tilde{E}_{hf}+E^{dyn}+E^{sta}$, where $\tilde{E}_{hf}$
is the Hartree-Fock energy obtained with natural orbitals. The new
procedure called NOF-MP2 scales formally as $O(M^{5})$ ($M$: number
of basis functions), and is applied successfully to the homolytic
dissociation of a selected set of diatomic molecules, paradigmatic
cases of near-degeneracy effects. Its size-consistency has been numerically
demonstrated for singlets. The values obtained are in good agreement
with the experimental data.
\end{abstract}
\maketitle
In electronic structure theory, accurate solutions require a balanced
treatment of both static (non-dynamic) and dynamic correlation. Static
correlation generally implies a limited number of nearby delocalized
orbitals with significant fractional occupations. Conversely, dynamic
correlation involves a large number of orbitals and configurations,
each with a small weight. 

Nowadays, it is necessary to resort to multi-reference methods for
correctly handling both types of correlation, however, these techniques
are often expensive and demand prior knowledge of the system. On the
other hand, single-reference correlation methods are well-established
for dynamic correlation, but are unsatisfactory for systems with static
correlation. The aim of this work is to propose a single-reference
method capable of achieving both dynamic and static correlation even
for those systems with significant multiconfigurational character.

In our approach, a natural orbital functional (NOF) \cite{Piris2007s}
is firstly used for capturing all static correlation effects. Then,
the total energy is approximated as $\tilde{E}_{hf}+E^{dyn}+E^{sta}$,
where $\tilde{E}_{hf}$ is the Hartree-Fock energy obtained with corresponding
natural orbitals (NOs). The dynamic energy correction ($E^{dyn}$)
is derived from a properly modified second-order Møller\textendash Plesset
perturbation theory (MP2) \cite{Moller1934}, while the non-dynamic
correction ($E^{sta}$) is obtained from the pure static component
of the new NOF. Let's start with the NOF.

In NOF theory, the spectral decomposition of the one-particle reduced
density matrix ($\Gamma={\displaystyle {\textstyle \sum}_{i}n_{i}}\left|\phi_{i}\right\rangle \left\langle \phi_{i}\right|$)
is used to approximate the electronic energy in terms of the NOs and
their occupation numbers (ONs), namely,
\begin{equation}
E=\sum\limits _{i}n_{i}\mathcal{H}_{ii}+\sum\limits _{ijkl}D[n_{i},n_{j},n_{k},n_{l}]<kl|ij>\label{ENOF}
\end{equation}
Here, $\mathcal{H}_{ii}$ denotes the diagonal elements of the core-Hamiltonian,
$<kl|ij>$ are the matrix elements of the two-particle interaction,
and $D[n_{i},n_{j},n_{k},n_{l}]$ represents the reconstructed two-particle
reduced density matrix (2-RDM) from the ONs.

It is noteworthy that the resulting approximate functional $E\left[\left\{ n_{i},\phi_{i}\right\} \right]$
can solely be implicitly dependent on $\Gamma$ \cite{Donnelly1979}
since the Gilbert's theorem \cite{Gilbert1975} on the existence of
the explicit functional $E\left[\Gamma\right]$ is valid only for
the exact ground-state energy. In this vein, NOs are the orbitals
that diagonalize the one-matrix corresponding to an approximate energy
that still depends on the 2-RDM \cite{Donnelly1979}. Consequently,
the energy is not invariant with respect to a unitary transformation
of the orbitals, and it is more appropriate to speak of a NOF rather
than a functional $E\left[\Gamma\right]$. A detailed account of the
state of the art of the NOF theory can be found elsewhere \cite{Piris2014a,Pernal2016}.

The construction of a\textit{ N}-representable functional given by
(\ref{ENOF}), i.e., derived from an antisymmetric \textit{N}-particle
wavefunction \cite{Coleman1963}, is obviously related to the \textit{N}
representability problem of the 2-RDM. The use of the (2,2)-positivity
\textit{N}-representability conditions \cite{Mazziotti2012} for generating
a reconstruction functional was proposed in reference \cite{Piris2006}.
This particular reconstruction is based on the introduction of two
auxiliary matrices $\mathbf{\triangle}$ and $\Pi$ expressed in terms
of the ONs to reconstruct the cumulant part of the 2-RDM \cite{Mazziotti1998}.
In a spin-restricted formulation, the energy functional for singlet
states reads as
\begin{equation}
\begin{array}{c}
E=2\sum\limits _{p}n_{p}\mathcal{H}_{pp}+\sum\limits _{qp}\Pi_{qp}\mathcal{L}_{pq}\qquad\qquad\\
+\sum\limits _{qp}\left(n_{q}n_{p}-\Delta_{qp}\right)\left(2\mathcal{J}_{pq}-\mathcal{K}_{pq}\right)
\end{array}\label{PNOF}
\end{equation}
where $\mathcal{J}_{pq}$, $\mathcal{K}_{pq}$, and $\mathcal{L}_{pq}$
are the direct, exchange, and exchange-time-inversion integrals \cite{Piris1999}.
Appropriate forms of matrices $\Delta$ and $\Pi$ lead to different
implementations known in the literature as PNOFi (i=1-6) \cite{Piris2014a}.

The conservation of the total spin allows to derive the diagonal elements
$\Delta_{pp}=n_{p}^{2}$ and $\Pi_{pp}=n_{p}$ \cite{Piris2009}.
The $N$-representability D and Q conditions of the 2-RDM impose the
following inequalities on the off-diagonal elements of $\Delta$ \cite{Piris2006},
\begin{equation}
\begin{array}{c}
\Delta_{qp}\leq n_{q}n_{p}\end{array},\qquad\Delta_{qp}\leq h_{q}h_{p}\label{DQ_cond}
\end{equation}
while to fulfill the G condition, the elements of the $\Pi$-matrix
must satisfy the constraint \cite{Piris2010a}
\begin{equation}
\Pi_{qp}^{2}\leq\left(n_{q}h_{p}+\Delta_{qp}\right)\left(h_{q}n_{p}+\Delta_{qp}\right)\label{G_cond}
\end{equation}
where $h_{p}$ denotes the hole $1-n_{p}$. Notice that for singlets
the total hole for a given spatial orbital $p$ is $2h_{p}$.

Let's divide the orbital space $\Omega$ into\textit{ $N/2$} mutually
disjoint subspaces $\Omega{}_{g}$, so each orbital belongs only to
one subspace. Consider each subspace contains one orbital $g$ below
the Fermi level (\textit{$N/2$}), and $N_{g}$ orbitals above it,
which is reflected in additional sum rules for the ONs:
\begin{equation}
\sum_{p\in\Omega_{g}}n_{p}=1;\quad g=1,2,\ldots,N/2\label{sumrule_n}
\end{equation}
Taking into account the spin, each subspace contains solely an electron
pair, and the normalization condition for $\Gamma$ ($2\sum_{p}n_{p}=N$)
is automatically fulfilled. It is important to note that orbitals
satisfying the pairing conditions (\ref{sumrule_n}) are not required
to remain fixed throughout the orbital optimization process \cite{Piris2009a}.

The simplest way to comply with all constraints leads to an independent
pair model (PNOF5) \cite{Piris2011,Piris2013e}:
\begin{equation}
\begin{array}{c}
\Delta_{qp}=n_{p}^{2}\delta_{qp}+n_{q}n_{p}\left(1-\delta_{qp}\right)\delta_{q\Omega_{g}}\delta_{p\Omega_{g}}\\
\\
\Pi_{qp}=n_{p}\delta_{qp}+\Pi_{qp}^{g}\left(1-\delta_{qp}\right)\delta_{q\Omega_{g}}\delta_{p\Omega_{g}}\\
\\
\Pi_{qp}^{g}=\left\{ \begin{array}{cc}
-\sqrt{n_{q}n_{p}}\,, & p=g\textrm{ or }q=g\\
+\sqrt{n_{q}n_{p}}\,, & p,q>N/2
\end{array}\right.\\
\\
\delta_{q\Omega_{g}}=\begin{cases}
1, & q\in\Omega_{g}\\
0, & q\notin\Omega_{g}
\end{cases}
\end{array}\label{PNOF5}
\end{equation}

Interestingly, an antisymmetrized product of strongly orthogonal geminals
(APSG) with the expansion coefficients explicitly expressed by the
ONs also leads to PNOF5 \cite{Pernal2013,Piris2013e}. This demonstrates
that the functional is strictly $N$-representable. In addition, PNOF5
is size-extensive and size-consistent, inherent properties to singlet-type
APSG wavefunctions.

To go beyond the independent-pair approximation, let's maintain $\mathrm{\Delta_{\mathit{qp}}=0}$
and consider nonzero the $\Pi$-elements between orbitals belonging
to different subspaces. From Eq. (\ref{G_cond}), note that provided
the $\Delta_{qp}$ vanishes, $\left|\Pi_{qp}\right|\leq\Phi_{q}\Phi_{p}$
with $\Phi_{q}=\sqrt{n_{q}h_{q}}$. Assuming the equality, one can
generalize the sign convention (\ref{PNOF5}), namely $\Pi_{qp}^{\Phi}=\Phi_{q}\Phi_{p}$
if $q,p>N/2,$ and $\Pi_{qp}^{\Phi}=-\Phi_{q}\Phi_{p}$, otherwise.
Thus, the energy (\ref{PNOF}) becomes
\begin{equation}
\begin{array}{c}
E=\sum\limits _{g=1}^{N/2}E_{g}+\sum\limits _{f\neq g}^{N/2}E_{fg}\\
\\
E_{g}=\sum\limits _{p\in\Omega_{g}}n_{p}\left(2\mathcal{H}_{pp}+\mathcal{J}_{pp}\right)+\sum\limits _{q,p\in\Omega_{g},q\neq p}\Pi_{qp}^{g}\mathcal{L}_{pq}\\
\\
E_{fg}=\sum\limits _{p\in\Omega_{f}}\sum\limits _{q\in\Omega_{g}}\left[n_{q}n_{p}\left(2\mathcal{J}_{pq}-\mathcal{K}_{pq}\right)+\Pi_{qp}^{\Phi}\mathcal{L}_{pq}\right]
\end{array}\label{PNOF7}
\end{equation}
This new approach will henceforth refer to as PNOF7. The first term
of the Eq.(\ref{PNOF7}) is the sum of the pair energies described
accurately by the two-electron functional $E^{g}$. In the second
term, $E^{fg}$ correlates the motion of the electrons in different
pairs with parallel and opposite spins. It is clear that the main
weaknesses of the approach (\ref{PNOF7}) is the absence of the interpair
dynamic electron correlation since $\Pi_{qp}^{\Phi}$ has significant
values only when the ONs differ substantially from 1 and 0. Consequently,
PNOF7 is expected to be able to recover the complete intra-pair, but
only the static inter-pair correlation.

\begin{figure}[b]
\caption{\label{fig:PECs}Potential Energy Curves (cc-pVTZ)}
\centering{}\includegraphics[scale=0.26]{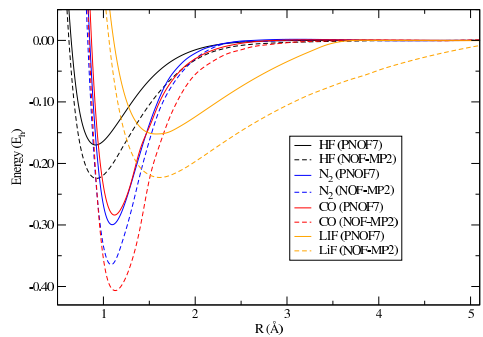}
\end{figure}

The performance of the PNOF7 has been tested by the dissociation of
a dozen diatomic molecules. The selected systems comprise different
types of bonding, which span a wide range of values for binding energies
and bond lengths. However, in all cases the correct dissociation limit
implies an homolytic cleavage of the bond with high degree of degeneracy
effects. For simplicity, we consider $N_{g}$ equal to a fixed number
that corresponds to the maximum value allowed by the basis set used

\begin{table}[h]
\caption{\label{tab:ComparisonPNOF7}Comparison of R$_{e}$ ($\textrm{\AA}$),
D$_{e}$ (kcal/mol), and $\omega_{e}$ (cm$^{-1}$) calculated at
the PNOF7/cc-pVTZ level of theory with the experimental values. $^{(a)}$
aug-cc-pVTZ was used.}
\centering{}%
\begin{tabular}{llccccccccccc}
\hline 
Molecule &  & R$_{e}$ &  & $\mathrm{R_{\mathit{e}}^{\mathit{exp}}}$ & \quad{} & D$_{e}$ &  & $\mathrm{D_{\mathit{e}}^{\mathit{exp}}}$ & \quad{} & $\omega_{e}$ &  & $\mathrm{\omega_{\mathit{e}}^{\mathit{exp}}}$\tabularnewline
\hline 
H$_{2}$ &  & 0.743 &  & 0.743 &  & 108.6 &  & 109.5 &  & 4404 &  & 4401\tabularnewline
LiH &  & 1.604 &  & 1.595 &  & \enskip{}56.1 &  & \enskip{}58.0 &  & 1404 &  & 1406\tabularnewline
Li$_{2}$ &  & 2.667 &  & 2.673 &  & \enskip{}23.3 &  & \enskip{}24.4 &  & \enskip{}330 &  & \enskip{}351\tabularnewline
BH &  & 1.232 &  & 1.232 &  & \enskip{}75.7 &  & \enskip{}81.5 &  & 2370 &  & 2367 \tabularnewline
OH$^{-(a)}$ &  & 0.966 &  & 0.964 &  & \enskip{}87.0 &  & - &  & 3010  &  & 3770\tabularnewline
HF &  & 0.915 &  & 0.917 &  & 106.7 &  & 141.1 &  & 4139 &  & 4138\tabularnewline
LiF &  & 1.576 &  & 1.564  &  & \enskip{}95.4 &  & 139.0 &  & \enskip{}668 &  & \enskip{}911\tabularnewline
N$_{2}$ &  & 1.097 &  & 1.098  &  & 188.9 &  & 228.3 &  & 2290 &  & 2359\tabularnewline
CN$^{-(a)}$ &  & 1.186 &  & 1.177 &  & 212.0 &  & 240.7 &  & 1999 &  & 2035\tabularnewline
CO &  & 1.120 &  & 1.128  &  & 178.1 &  & 259.3 &  & 2316 &  & 2170\tabularnewline
NO$^{+}$ &  & 1.056 &  & 1.063 &  & 179.9 &  & - &  & 2412 &  & 2377\tabularnewline
F$_{2}$ &  & 1.579 &  & 1.412  &  & \enskip{}\enskip{}2.6 &  & \enskip{}39.2 &  & \enskip{}422 &  & \enskip{}917\tabularnewline
\hline 
\end{tabular}
\end{table}

Representative potential energy curves (PECs) of these molecules are
depicted in Fig.\ref{fig:PECs} (see supplementary material for absolute
energies). PNOF7 produces qualitatively correct PECs with right dissociation
limits for all cases, even in the case of the highest degeneracy ($\mathrm{N}_{2}$).
In Table \ref{tab:ComparisonPNOF7}, selected electronic properties,
including equilibrium bond lengths (R$_{e}$), dissociation energies
(D$_{e}$), and harmonic vibrational frequencies ($\omega_{e}$) can
be found. In this work, the experimental bond lengths and spectroscopic
data reported are taken from the National Institute of Standards and
Technology (NIST) Database \cite{nist}, whereas the experimental
dissociation energies result from a combination of Refs. \cite{nist}
and \cite{Chase1998}. The correlation-consistent valence triple-$\zeta$
basis set (cc-pVTZ) developed by Dunning \cite{Dunning1989} was used
throughout, except for the anionic species (OH$^{-}$ and CN$^{-}$)
where the augmented basis set (aug-cc-pVTZ) was used.

Table \ref{tab:ComparisonPNOF7} shows that the results are in good
agreement with the experiment for the smaller diatomics, for which
the electron correlation effect is almost entirely intrapair. When
the number of pairs increases, the theoretical values deteriorate
especially for the dissociation energies. This is related to a better
description of the asymptotic region with respect to the equilibrium
where the dynamic correlation prevails. It is therefore mandatory
to add the inter-pair dynamic electron correlation to improve these
results.

The second-order Møller\textendash Plesset \cite{Moller1934} perturbation
theory (MP2) is the simplest and cheapest way of properly incorporating
dynamic electron correlation effects. The Hartree-Fock (HF) wavefunction
is taken as the starting point in MP2, so let's consider an Slater
determinant formed by the NOs as the zeroth-order wavefunction, and
define the zeroth-order Hamiltonian $\mathrm{\hat{H}^{(0)}}$ by the
expansion $\sum_{i}\varepsilon_{i}\left|\phi_{i}\right\rangle \left\langle \phi_{i}\right|$.
Here, $\varepsilon_{i}$ is the $i$-th diagonal element of the Fock
matrix ($\mathcal{F}$) in the NO representation. The first-order
energy correction leads to an energy ($\tilde{E}_{hf}$) that differs
from the true HF energy since NOs are used instead of the canonical
HF orbitals. Besides, the Fock matrix is no longer diagonal, therefore
single excitations in addition to doubles contribute to the MP2 energy
correction, namely,
\begin{equation}
\begin{array}{c}
E^{\left(2\right)}=2\sum\limits _{g=1}^{N/2}\,\sum\limits _{p>N/2}^{M}\,{\displaystyle \frac{\left|\mathcal{F}_{pg}\right|^{2}}{\varepsilon_{g}-\varepsilon_{p}}}+\sum\limits _{g,f=1}^{N/2}\quad\\
\\
\sum\limits _{p,q>N/2}^{M}{\displaystyle \frac{\left\langle gf\right|\left.pq\right\rangle \left[2\left\langle pq\right|\left.gf\right\rangle -\left\langle pq\right|\left.fg\right\rangle \right]}{\varepsilon_{g}+\varepsilon_{f}-\varepsilon_{p}-\varepsilon_{q}}}
\end{array}\label{E2}
\end{equation}
where $M$ is the number of basis functions.

\begin{figure}
\caption{\label{fig:PEC-H2}Potential energy curves for H$_{2}$.}
\centering{}\includegraphics[scale=0.26]{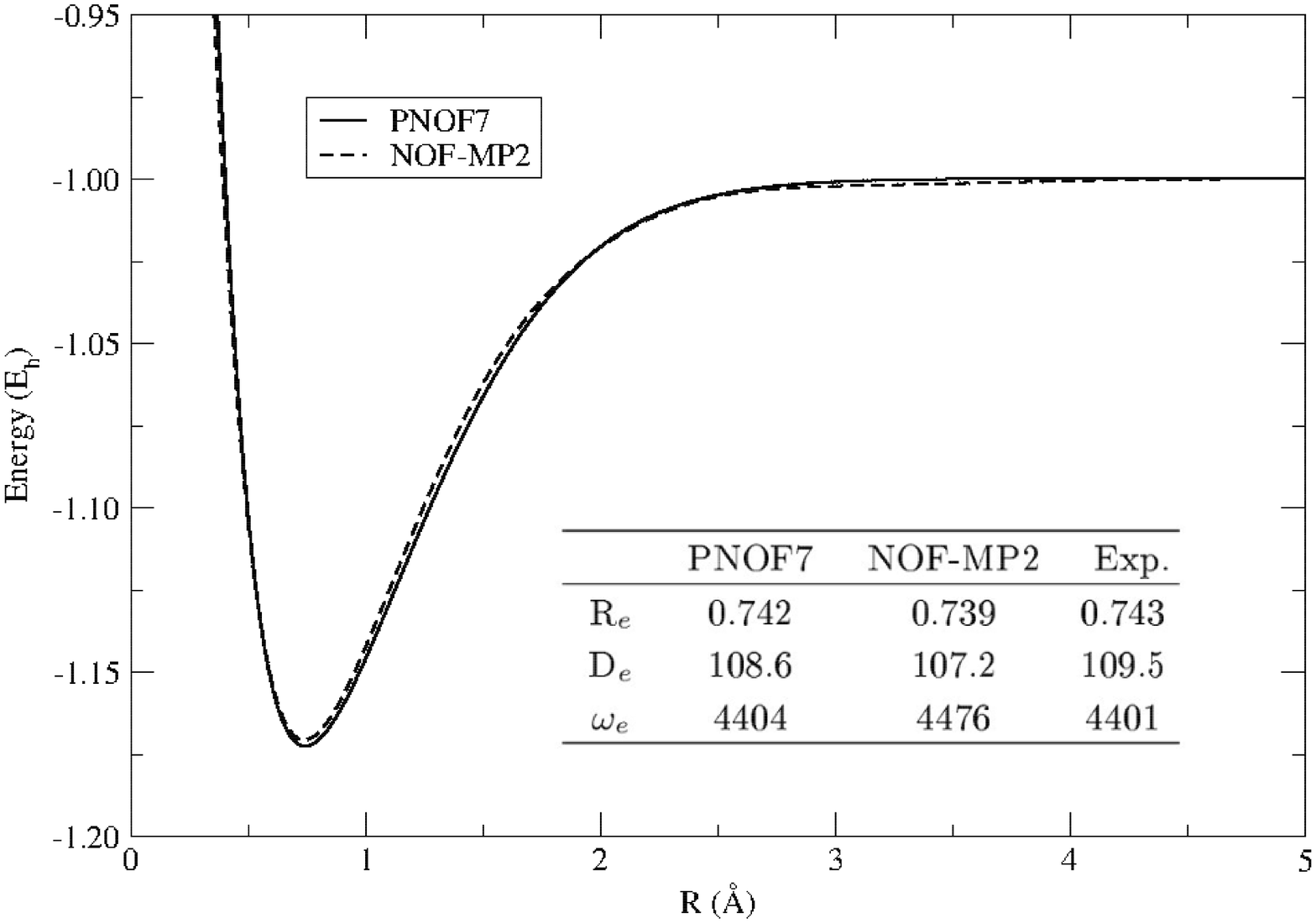}
\end{figure}

In general, MP2 lacks non-dynamic correlation, which is well recovered
by PNOF7, but we cannot simply add these contributions since double
counting occurs. With this in mind, new dynamic ($E^{dyn}$) and static
($E^{sta}$) energy functionals have to be defined from the MP2 and
PNOF7, respectively, so that the total energy of the system will be
given by
\begin{equation}
E=\tilde{E}_{hf}+E^{corr}=\tilde{E}_{hf}+E^{sta}+E^{dyn}\label{Etotal}
\end{equation}

Henceforth, the energy obtained with the Eq. (\ref{Etotal}) is called
NOF-MP2. From Eq. (\ref{PNOF7}), it is evident that we must differentiate
between intra- and inter-pair contributions for both functionals.
In accordance, one has
\begin{equation}
\begin{array}{c}
E_{intra}^{corr}=\sum\limits _{g=1}^{N/2}\left(E_{g}^{sta}+E_{g}^{dyn}\right)\\
E_{inter}^{corr}=\sum\limits _{f\neq g}^{N/2}\left(E_{fg}^{sta}+E_{fg}^{dyn}\right)
\end{array}
\end{equation}
hence $E^{corr}=E_{intra}^{corr}+E_{inter}^{corr}$ as well. To avoid
double counting, we are going to consider the amount of static electron
correlation in each orbital as a function of its occupancy:
\begin{equation}
\Lambda_{p}=1-\left|1-2n_{p}\right|\label{fs}
\end{equation}
Note that $\Lambda_{p}$ goes from zero for empty or fully occupied
orbitals to one if the orbital is half occupied. Using this function,
let's define the static and dynamic $g$-th intra-pair electron correlation
energies as
\begin{equation}
\begin{array}{c}
E_{g}^{sta}=\sum\limits _{q\neq p}\sqrt{\Lambda_{q}\Lambda_{p}}\,\Pi_{qp}^{g}\mathcal{\,L}_{pq}\\
\\
E_{g}^{dyn}=2C_{g}\sum\limits _{p>N/2}^{M}C_{p}{\displaystyle \frac{\left|\mathcal{F}_{pg}\right|^{2}}{\varepsilon_{g}-\varepsilon_{p}}}\\
\\
+\,C_{g}^{2}\sum\limits _{p,q>N/2}^{M}{\displaystyle C_{p}C_{q}\frac{\left\langle gg\right|\left.pq\right\rangle \left\langle pq\right|\left.gg\right\rangle }{2\varepsilon_{g}-\varepsilon_{p}-\varepsilon_{q}}}
\end{array}\label{Eintra}
\end{equation}

where $q,p\in\Omega_{g}$ and $C_{p}=1-\Lambda_{p}^{2}$. The PECs
for the archetypal 2-electron singlet, H$_{2}$, are depicted in Fig.
\ref{fig:PEC-H2}. It is remarkable the excellent agreement between
the results obtained with the new intra-pair energy functionals given
by Ec. (\ref{Eintra}) and those of PNOF7, which in this case is practically
exact \cite{Hirschfelder1959}.

Taking into account the square root that already appears in the definition
of the $\Phi$ magnitudes, we can similarly introduce the following
functionals for the $fg$-th inter-pair static and dynamic correlation
energies:
\begin{equation}
\begin{array}{c}
E_{fg}^{sta}=\sum\limits _{p\in\Omega_{f}}\sum\limits _{q\in\Omega_{g}}4\Phi_{p}\Phi_{q}\,\Pi_{qp}^{\Phi}\mathcal{\,L}_{pq}\\
\\
E_{fg}^{dyn}=2\sum\limits _{p>N/2}^{M}{\displaystyle \delta_{p\Omega_{f}}C_{p}^{\Phi}\frac{\left|\mathcal{F}_{pg}\right|^{2}}{\varepsilon_{g}-\varepsilon_{p}}}+\sum\limits _{p,q>N/2}^{M}\delta_{p\Omega_{f}}\\
\\
C_{p}^{\Phi}\delta_{q\Omega_{g}}C_{q}^{\Phi}{\displaystyle \frac{\left\langle gf\right|\left.pq\right\rangle \left[2\left\langle pq\right|\left.gf\right\rangle -\left\langle pq\right|\left.fg\right\rangle \right]}{\varepsilon_{g}+\varepsilon_{f}-\varepsilon_{p}-\varepsilon_{q}}}
\end{array}\label{Einter}
\end{equation}
In Eq. (\ref{Einter}), $2\Phi_{p}$ plays the same role of $\sqrt{\Lambda_{p}}$
in Eq. (\ref{Eintra}), hence, $C_{p}^{\Phi}=1-4\Phi_{p}^{2}=1-4n_{p}h_{p}$.
Again, fully occupied and empty orbitals contribute nothing to static
correlation, this time inter-pair, whereas orbitals with half occupancies
yield a maximal contribution. The opposite occurs for dynamic correlation.
It is worth noting that $C^{\Phi}$ is not considered if the orbital
is below $N/2$.

\begin{table}
\caption{\label{tab:Comparison}Comparison of R$_{e}$ ($\textrm{\AA}$), D$_{e}$
(kcal/mol), and $\omega_{e}$ (cm$^{-1}$) calculated at the NOF-MP2/cc-pVTZ
level of theory with the experimental values. $^{(a)}$ aug-cc-pVTZ
was used.}
\centering{}%
\begin{tabular}{llccccccccccc}
\hline 
Molecule &  & R$_{e}$ &  & $\mathrm{R_{\mathit{e}}^{\mathit{exp}}}$ & \quad{} & D$_{e}$ &  & $\mathrm{D_{\mathit{e}}^{\mathit{exp}}}$ & \quad{} & $\omega_{e}$ &  & $\mathrm{\omega_{\mathit{e}}^{\mathit{exp}}}$\tabularnewline
\hline 
Be$_{2}$$^{(a)}$ &  & 2.303 &  & 2.460 &  & \enskip{}\enskip{}2.6 &  & \enskip{}\enskip{}2.7 &  & \enskip{}543 &  & -\tabularnewline
OH$^{-}{}^{(a)}$ &  & 0.967 &  & 0.964 &  & 121.6 &  & - &  & 3820  &  & 3770\tabularnewline
HF &  & 0.924 &  & 0.917 &  & 139.4 &  & 141.1 &  & 4151  &  & 4138\tabularnewline
LiF &  & 1.614 &  & 1.564  &  & 140.7 &  & 139.0 &  & \enskip{}955 &  & \enskip{}911\tabularnewline
N$_{2}$ &  & 1.084 &  & 1.098  &  & 224.2 &  & 228.3 &  & 2764 &  & 2359\tabularnewline
CN$^{-}{}^{(a)}$ &  & 1.180 &  & 1.177 &  & 238.6 &  & 240.7 &  & 1961  &  & 2035\tabularnewline
CO &  & 1.129 &  & 1.128  &  & 255.1 &  & 259.3 &  & 2092 &  & 2170\tabularnewline
NO$^{+}$ &  & 1.060 &  & 1.063 &  & 261.1 &  & - &  & 2403  &  & 2377\tabularnewline
F$_{2}$ &  & 1.397 &  & 1.412  &  & \enskip{}34.5 &  & \enskip{}39.2 &  & \enskip{}949 &  & \enskip{}917\tabularnewline
\hline 
\end{tabular}
\end{table}

Table \ref{tab:Comparison} collects the electronic properties previously
analyzed for the systems in which the inter-pair correlation becomes
important. The data reveals an outstanding improvement in the dissociation
energies, as well as a nice agreement of R$_{e}$ and $w_{e}$ with
the experimental marks. It is worth mentioning the case of F$_{2}$,
and the recovered correct order in the dissociation energies of the
N$_{2}$ and CO (see also Fig. \ref{fig:PECs}). 

The included Beryllium dimer requires special attention. PNOF7 predicts
a metastable minimum with a negative D$_{e}$, whereas NOF-MP2 recovers
sufficient dynamic correlation to be able of predicting a stable Be$_{2}$
molecule. The obtained equilibrium distance is still underestimated,
but the dissociation energy approaches the experimental value. For
weaker bonds, e.g. He$_{2}$, NOF-MP2 does not predict bound due to
a better description of the dissociated atoms with respect to the
equilibrium region. In these cases, neglecting static correlation
and using HF-MP2 leads to a binding PEC. The alternative is to include
higher-order perturbative corrections.

The size-consistency of the NOF-MP2, i.e. the ability of the method
to reproduce the additivity of the energy for a system composed of
independent subsystems, has been numerically addressed too. It has
been checked that total energies of spin-compensated dimers (He$_{2}$,
Be$_{2}$, and HeNe) at an internuclear separation of 100 $\textrm{\AA}$
differ from the double value of the total energies of the corresponding
atoms lesser than $10^{-5}$ Hartrees (< 0.01 kcal/mol).

Preliminary calculations on systems with more than two atoms confirm
that the results are promising. The absolute energies obtained with
the NOF-MP2 method improve over the PNOF7 values by recovering an
important part of the dynamic correlation and getting closer to the
values obtained by accurate wavefunction-based methods (see supplementary
material).

In summary, a new size-consistent method for singlet states has been
proposed that scales formally as $O(M^{5})$. The resulting working
formulas allow for static and dynamic correlation to be achieved in
one shot, as is the case in the standard single-reference perturbation
theory. Note that the NOF-MP2 method is not limited to PNOF7 NOs,
it can also be used with NOs obtained from an approximation able of
recovering non-dynamic electron correlation. In addition, the number
of orbitals involved in the optimization can be easily reduced by
establishing a cutoff in the value of the ONs, since the dynamic correlation
for which the orbitals with small ONs are responsible will be properly
recovered by $E^{dyn}$. With efficient approaches, based on recent
developments of NOF and MP2 theories, NOF-MP2 could become a valuable
tool for treating large systems with hundreds of atoms.

\selectlanguage{american}%
\textbf{Acknowledgments:} Financial support comes from MINECO (Ref.
CTQ2015-67608-P). The author thanks for the support provided by IZO-SGI
SGIker of UPV/EHU and European funding (ERDF and ESF).

\selectlanguage{english}%
%\bibliographystyle{apsrev}
%\addcontentsline{toc}{section}{\refname}\bibliography{manuscript}

\end{document}